\documentstyle[a4,11pt,epsf]{article}
\epsfverbosetrue
\newcommand{\beq}{\begin{equation}}
\newcommand{\eeq}{\end{equation}}   

 0

\begin{document}

\begin{titlepage}
 
\begin{flushright}
Liverpool Preprint: LTH 343\\


\end{flushright}

\vspace{5mm}
 
\begin{center}
{\Huge Logarithmic Corrections to Scaling in the Two\\ [3mm]
Dimensional $XY$--Model}\\[15mm]
{\bf R. Kenna\footnote{Supported by EU Human Capital 
and Mobility Scheme Project No. CHBI--CT94--1125} 
and A.C. Irving \\
~\\ 
DAMTP, University of Liverpool L69 3BX, England} \\[3mm]
December 1994
\end{center}
\begin{abstract}
By expressing thermodynamic functions in terms of the edge and density
of  Lee--Yang  zeroes, we relate the scaling behaviour of the specific
heat  to  that  of  the  zero  field  magnetic  susceptibility  in the 
thermodynamic   limit of the $XY$--model in two dimensions.   Assuming 
that  finite--size     scaling    holds, we show that the conventional 
Kosterlitz--Thouless    scaling predictions  for  these thermodynamic 
functions  are  not  mutually  compatable  unless they are modified by
multiplicative  logarithmic corrections. We identify these logarithmic 
corrections  analytically  in  the case  of  the  specific  heat   and 
numerically  in  the   case  of  the  susceptibility.  The  techniques 
presented here are general and can be used to  check the compatibility  
of scaling behaviour of odd and even thermodynamic functions in  other 
models too.
\end{abstract}
\end{titlepage}

\paragraph{The Kosterlitz--Thouless Scenario}
 

The  partition  function  for  the  $O(n)$ non--linear $\sigma$--model
defined on a  regular $d$--dimensional lattice $\Lambda$ with periodic 
boundary conditions in the presence  of  an  external  magnetic  field  
can be written as

\begin{equation}
 Z_L(\beta,h)
 =
 \frac{1}{\cal{N}}
 \sum_{\{{\vec{\sigma}}_x\}}{
                   e^{\beta S
                    + h {\vec{n}}\cdot{\vec{M}}
                   }
               }
\, ,
\label{pf}
\end{equation}
where ${\vec{n}}$ is a unit vector defining the direction of the external
magnetic field and $h$ is a scalar parameter representing its strength.
The summation is over all configurations open to the system.
The factor ${\cal{N}}$  ensures that $Z_L(0,0) =
1$ and 
\begin{displaymath}
 S
 =
 \sum_{x \in \Lambda}
 \sum_{\mu = 1}^{d}
 {\vec{\sigma}}_x\cdot
 {\vec{\sigma}}_{x+\mu}
\, ,         
 \quad \quad 
 {\vec{M}} =      
 \sum_{x \in \Lambda}
 {\vec{\sigma}}_x
\, .
\end{displaymath}
Here $L$ represents the linear extent of the  system, $ \beta=1/kT $ 
and ${\vec{\sigma}}_x$ is an $n$--component unit length spin at site 
$x \in \Lambda$.  
The case  $d=2$,  $n=2$ is the two dimensional $XY$--model 
or plane rotator model and is unusual in that it exhibits an
exponentional singularity.
Using an  approximate renormalization group approach, Kosterlitz  and  
Thouless \cite{KT} demonstrated the existence of a phase transition 
driven  by the condensation of vortices. 
The model remains critical  (thermodynamic functions diverge) for all
$\beta > \beta_c$ and the critical exponents are dependent on temperature.
In terms of the reduced temperature
\begin{displaymath}
  t = 1 - \frac{\beta}{\beta_c}
\, ,
\end{displaymath}
the scaling  behaviour of the correlation 
length, susceptibility and  the  specific  heat is given in \cite{KT} as
\begin{eqnarray}
 \xi_\infty(t) & \sim & e^{at^{-\nu}} \, , \label{ktxi} 
 \\
 \chi_\infty(t) & \sim & \xi_\infty^{2-\eta}\, , \label{ktchi}
 \\
 C_\infty(t) & \sim & \xi_\infty^{\tilde{\alpha}} + {\rm{constant}}\, ,
 \label{ktcv}   
\end{eqnarray}
where for  $t \rightarrow 0^+$,
$\nu = 1/2$, $\eta = 1/4$ and $\tilde{\alpha} =-d =-2$.
The purpose of this work is to argue that the latter two  scaling  formulae  are
incompatible as they stand. To this end, a method is presented by which  odd 
and even  thermodynamic  functions (the susceptibility and
specific heat) can 
be  related  and  expressed  in  terms  of  partition function zeroes. Using 
certain reasonable assumptions regarding finite--size scaling,  it  is shown 
that  there  have  to  exist   multiplicative   logarithmic   corrections
\footnote{The work of \cite{KT} contains implicitly
a prediction for logarithmic corrections but this seems not to have
been followed up quantitatively.} 
to (\ref{ktchi}) and (\ref{ktcv}). 
This  method  can be applied to any model. 

\paragraph{Partition Function Zeroes}
 
 
Lee and Yang \cite{LY} showed the connection 
between critical behaviour and partition function zeroes.
For a finite system these are strictly complex (non--real).
As $L\rightarrow\infty$ one expects the zeroes
to condense onto smooth curves. Zeroes in the plane of complex external
magnetic field $h$ are refered to as Lee--Yang zeroes. Lee and Yang further
showed that for certain systems these zeroes are in fact
restricted to the imaginary $h$ axis (the Lee--Yang theorem) \cite{LY}.
In the symmetric phase, $t>0$, they lie away from the real $h$-axis,
pinching it only as $t \rightarrow  0^+$ (in the thermodynamic limit).
For systems obeying the Lee--Yang theorem, such as the XY Model
\cite{DuNe75}, this pinching occurs at $h=0$
prohibiting analytic continuation from ${\rm{Re}}(h) < 0$ to
${\rm{Re}}(h) > 0 $.
This means that (for the finite--size system or in the thermodynamic limit) 
the partition function
is an entire function of $h$ all of whose zeroes are purely imaginary.
The zeroes in the complex fugacity plane ($z\equiv\exp{(h)}$) are distributed
on the unit circle. Writing these zeroes as
\begin{displaymath}
 z_j(\beta) = e^{i \theta_j (\beta)}\, ,
\end{displaymath}
the partition function is
\begin{equation}
 Z_L(\beta,h)
 =
 \rho_L(\beta,h) \prod_j{\left(z-e^{i\theta_j(\beta)}\right)}\, ,
\label{pfz}
\end{equation}
where $\rho_L$ is a non-vanishing function of $h$ related to the spectral
density (ignored in what follows since it contributes only to the 
regular  part  of  the free energy
in the thermodynamic limit).
The free energy is then 
\begin{equation}
 f_L(\beta,h) 
 = 
 L^{-d}
 \sum_j{ \ln{ (z-e^{  i\theta_j(\beta)} ) } }
\, .
\label{free}
\end{equation}
Define the density of zeroes to be \cite{KeLa94,Salmhofer}
\begin{equation}
 g_L(\beta,\theta )
 =
 L^{-d}
 \sum_j{\delta(\theta-\theta_j(\beta))}
 =
 \frac{\partial G_L(\beta,\theta)}{\partial \theta}
\, .
\label{cum}
\end{equation}
Here $G_L(\beta,\theta)$ is the cumulative density of zeroes
and monotonically increases from 
$G_L(\beta,0)=0$ to $G_L(\beta,\pi)=1/2$.
The distribution of zeroes
is symmetric about the real $h$-- (or $z$--) axis 
and so
\begin{equation}
 g_L(\beta,-\theta) = g_L(\beta, \theta)
\, .
\label{symm}
\end{equation}
In terms of the density of zeroes, (\ref{free}) becomes
\begin{equation}
 f_L( \beta,h ) = 
 \frac{1}{2}(h + \ln{2})
 +
  \int_{0}^{\pi}{   
                   \ln{ \left( \cosh{h} - \cos{\theta}  \right)
                      } 
                   d G_L (\beta,\theta) 
                  } 
\, ,
\label{eq:fLint}
\end{equation}
where (\ref{symm}) has been used.

In the high temperature phase there exists a region around $\theta=0$
which is free from Lee--Yang zeroes \cite{LY}.
We define the corresponding Yang--Lee edge $\theta_{{\rm{YL}}}(\beta)$ by 
\begin{displaymath}
 g(\beta,\theta)=0 {\rm{~ ~ for ~ ~}} 
 -\theta_{{\rm{YL}}}(\beta) < \theta < \theta_{{\rm{YL}}}(\beta)
\, .
\end{displaymath}
The integral in (\ref{eq:fLint}) can therefore be taken to begin at 
$\theta_{{\rm{YL}}}(\beta)$.
Salmhofer has proved the existence of a unique density of zeroes
in the thermodynamic limit \cite{Salmhofer}.
In this limit, integrating (\ref{eq:fLint}) by parts
and expanding the trigonometric functions gives for the singular
part of the free energy \cite{KeLa94,Abe,SuzukiLY}
\begin{equation}
 f_{\infty}(\beta,h)
 = 
 -
 2 \int_{\theta_{{\rm{YL}}}(\beta)}^{\pi}
 \frac{\theta}{h^2 + \theta^2}
 G_\infty(\beta,\theta)
 d\theta
\, .
\label{free1}
\end{equation}
The  magnetic   susceptibility    $\chi_\infty$   is  given by the second 
derivative of the  free  energy  with  respect  to  $h$. 
Following \cite{KeLa94,Abe,SuzukiLY}, this leads to 
\begin{equation}
 G_\infty(\beta,\theta)
 =
 \chi_\infty (\beta)
 {( \theta_{{\rm{YL}}}(\beta)) }^2
 \Phi \left( \frac{\theta}{\theta_{{\rm{YL}}}(\beta)} \right)
\, ,
\label{twoPhi}
\end{equation}
$\Phi (x)$ being some function of $x$ such that
$\Phi(\mid x\mid\leq 1)=0$. 
From (\ref{free1}) and the fact that
$G\left(\theta_{{\rm{YL}}},\beta\right)=0$, one gets the specific heat
\begin{equation}
C_\infty(\beta) =  \left.
 \frac{\partial^2f_\infty(\beta,h)}{\partial \beta^2}
 \right|_{h=0}
=
 -2\int_{\theta_{{\rm{YL}}}(\beta)}^\pi
 \theta^{-1}
 \frac{d^2G_\infty\left(\theta,\beta \right)}{d\beta^2}
 d\theta
 \, .
\label{directerthanAbe}
\end{equation}
The above formulae are quite general and hold for any model provided
it obeys the Lee--Yang theorem.

To proceed further we insert the (model--specific) critical behaviour. 
Instead of 
(\ref{ktchi}) and (\ref{ktcv}), assume now the following 
modified critical behaviour for the singular parts of the zero field
susceptibility and the specific heat for
$t$ \raisebox{-.75ex}{ {\small \shortstack{$>$ \\ $\sim$}} } $0$.
\begin{equation}
 \chi_\infty \sim  \xi_\infty^{2-\eta} t^r
\, ,
\quad \quad 
  C_\infty   \sim  \xi^{\tilde{\alpha}} t^q
\, .
\end{equation}
Similarly, assume that the leading scaling behaviour of the Yang--Lee 
edge in terms of the reduced temperature is 
\begin{equation}
 \theta_{{\rm{YL}}}(t) \sim \xi^{\lambda} t^p
\, .
\label{singedge}
\end{equation}
For $t$ \raisebox{-.75ex}{ {\small \shortstack{$>$ \\ $\sim$}} } $0$
the critical indices are independent of $t$ \cite{KT}.
The scaling behaviour of $C_\infty$ can be related to that of $\chi_\infty$
and $\theta_{{\rm{YL}}}$ through (\ref{twoPhi}) and (\ref{directerthanAbe}).
These give
\begin{displaymath}
 {\xi(t)}^{{\tilde{\alpha}}}
 t^{q}
 \propto
 {\xi(t)}^{ 2 - \eta  + 2 \lambda }
 t^{2p+r -2 -2\nu}
\, .
\end{displaymath}
Therefore,
\begin{eqnarray}
 \lambda 
 & = &
 \frac{1}{2}
 ( {\tilde{\alpha}} -2 + \eta )\, ,
\\
 p & = & \frac{1}{2}(q-r) + 1 + \nu
\, .
\end{eqnarray}

\paragraph{Finite--Size Scaling}
 

The finite--size scaling (FSS) hypothesis, first formulated in 1972
by Fisher \cite{Fi72},  is a relationship
between the scaling behaviour of thermodynamic quantities in the infinite 
volume limit and the size dependence of their finite volume counterparts.
The general statement of FSS, which is expected to hold in all dimensions
including the upper critical one \cite{KeLa93} is that if
$P_L(t)$ is the value of some thermodynamic quantity $P$ at reduced 
temperature $t$ measured on a system of linear extent $L$, then
\begin{equation}
 \frac{P_L(0)}{P_\infty (t)}
 =
 {\cal{F}}_P
 \left( 
        \frac{\xi_L(0)}{\xi_\infty (t)}
 \right)
\, ,
\label{FSS}
\end{equation}
where $\xi_L (t)$ is the correlation length of the finite--size system.
Luck has shown that, for the $XY$--model in two dimensions,
$\xi_L(0)$ is proportional to $L$ \cite{Lu82}.
Fixing the scaling variable, one has
\begin{displaymath}
 \xi_\infty (t) 
 \sim
 x^{-1} L
\, ,
\end{displaymath}
and from (\ref{ktxi})
\begin{displaymath}
 t \sim (\ln{L})^{-\frac{1}{\nu}}
\, ,
\end{displaymath}
for large enough $L$.
Therefore FSS for the susceptibility and the Yang--Lee edge is
\begin{eqnarray}
 \chi_L (0) & \sim & L^{2 - \eta} (\ln{L})^{-\frac{r}{\nu}}\, ,
\label{chifss}
 \\
  \theta_1(0)   &  \sim & L^{\lambda}  (\ln{L})^{- \frac{p}{\nu}}
\label{edgefss}
\, .
\end{eqnarray}
For the finite size system it is convenient to consider the zeroes
in the complex $h$--plane. The singular part of the free energy
corresponds to
\begin{equation}
 f_L(t,h)
 =
 L^{-d}
 \sum_j{\ln{\left( h - i \theta_j(t)  \right)}}
\end{equation}
where $\theta_{1}=\theta_{YL}$.
The
(reduced) magnetic susceptibility is therefore
\begin{equation}
 \chi_L (t) 
 =
 \left.
    \frac{\partial^2 f_L}{\partial h^2}
 \right|_{h=0}
 =
 -
 L^d
 \sum_j{
          \frac{1}{\theta_j^2}
}
\, .
\end{equation}
The lowest lying zeroes are expected to scale in the same way 
\cite{IPZ} so that
\begin{equation}
 \chi_L (t) \propto - L^{-d} \theta_1(t)^{-2}
\, .
\end{equation}
From (\ref{chifss}) and (\ref{edgefss}) we have
\begin{equation}
 L^{2 - \eta}
 (\ln{L})^{-\frac{r}{\nu}}
 \sim
 L^{-d - 2 \lambda}
 (\ln{L})^{ \frac{2p}{\nu} }         
\, ,
\label{fssf}
\end{equation}
and conclude
\begin{equation}
 {\tilde{\alpha}}  =  - d\, ,\qquad  q  = - 2 (1+ \nu) 
\, .
\end{equation}
Thus the scaling behaviour of the singular part of the specific heat 
indeed exhibits multiplicative logarithmic corrections.  
Accepting the KT predictions $\nu=1/2$, $\eta=1/4$ 
for $t$ \raisebox{-.75ex}{ {\small \shortstack{$>$ \\ $\sim$}} } $0$,
the leading critical exponent for the Yang--Lee edge $\lambda$ and
the correction exponents $q$ and $p$ are
\begin{equation}
 \lambda = - \frac{15}{8}
\, ,
 \quad \quad 
 q= - 3
\, ,
 \quad \quad
 p = - \frac{r}{2}
\, .
\label{pqr}
\end{equation}
The above analytic considerations have yielded no information on
the odd correction exponent $r$. 
The original
renormalisation group analysis of Kosterlitz and Thouless \cite{KT}, in fact,
implicitly contained the prediction 
\beq
r=-1/16
\label{eq:r16}
\eeq
as noted by \cite{BuCo92}. Subsequent analysis have concentrated on the
$r=0$ form of the scaling behaviour (\ref{chifss})
and the verification that $\eta(\beta_c)=1/4$.
Allton and Hamer \cite{AlHa88} have conjectured that the
deviation of their determination of $\eta$ from $1/4$ might
be due to logarithmic corrections.

The study of the full scaling form and the numerical determination of $r$
is the subject of what follows.

\paragraph{Numerical Procedures}
 

Numerical methods were used to test the scaling picture of the last
section.
Specifically, we analysed the FSS behaviour of the Yang--Lee edge so 
as to test the prediction (\ref{edgefss}), (\ref{pqr})
\begin{equation}
 \theta_1(\beta_c) \sim L^\lambda 
 (\ln{L})^r
\, .
\label{goal}
\end{equation}
In particular, we sought to confirm $\lambda = 15/8$ for the leading
behaviour and to check if $r\neq 0$.

An algorithm based on that of Wolff \cite{Wolff} 
was used to simulate the $XY$--model at zero magnetic field
($h=0$) on square lattices of sizes $L=32,64,128$  and  $256$ . 
The values of $\beta$ at which the simulations took place ($\beta_o$)
were evenly spaced in steps of $0.02$ between $1.00$ and $1.20$.
At each
simulation point, $50,000$ measurements were made from each of two
separate starts. The second runs were used only to verify 
equilibration and to improve our understanding of the statistical 
errors. The latter were
estimated using bootstrap as described below. 
The autocorrelation time of the susceptibility was also monitored. 
Histogram reweighting \cite{histo}
was used to enable extrapolation between $\beta_o$
values.

The determination of $\beta_c$ is inextricably bound up with
the assumed critical scaling behaviour. 
For the present purposes, we adopted a two-stage
strategy: we used estimates of $\beta_c$ from our own and other
analyses which had assumed {\it no} logarithmic corrections then
tested the stability of these conclusions with respect both to uncertainties
in $\beta_c$ and to the form of the scaling asumption.

One simple estimate was based on assumed conventional ($r=0$) scaling
of the susceptibility (\ref{ktchi}). A generalised \cite{HI}
Roomany-Wyld approximant \cite{RoWy} was used to give estimates of
the corresponding renormalisation group beta function for each
pair of lattice sizes related by factor of two scaling: 
\beq
\beta^{RW}_L=\bigl({2\over{\beta^2}}\bigr)
{
{\eta-2+
2r\ln( {{\ln 2L}\over{\ln L}})/\ln 2 +
\ln(\chi_{2L}/\chi_L)/\ln 2
}
\over
{{\partial\over{\partial\beta}}[\chi_{2L}\chi_L]}
}\, .
\label{eq:RWA}
\eeq

A preliminary estimate, using $\eta=1/4$ and $r=0$, leads to 
to $\beta_c=1.11(1)$ which is at the lower
end of the range of published estimates 
\cite{BuCo92},
\cite{many},
\cite{HaMa94}.
These span $1.11$ to $1.13$. The most recent high precision result
is $1.1197(5)$ \cite{HaMa94}.

\paragraph{Determination of Lee--Yang Zeroes}
 
When the external field
is complex ($h = h_r + i h_i $), 
the partition function (\ref{pf}) can be rewritten
\begin{equation}
  Z_L(\beta,h_r + i h_i)
  =
  {\rm{Re}} Z_L(\beta,h_r + i h_i)
  + i
  {\rm{Im}} Z_L(\beta,h_r + i h_i)
\, ,
\end{equation}
where
\begin{eqnarray}
 {\rm{Re}} Z_L(\beta,h_r + i h_i)
 & = &
 \frac{1}{\cal{N}}
 \sum_{\{{\vec{\sigma}}_x\}}{e^{\beta S + h_r M} \cos{(h_i M)}}
\, ,
\nonumber
\\
 ~
 & = &
 Z_L(\beta,h_r)
 \langle
 \cos{(h_i M)}
 \rangle_{\beta,h_r}
\label{ReZ}
\, ,
\end{eqnarray}
and
\begin{eqnarray}
 {\rm{Im}} Z_L(\beta,h_r + i h_i)
 & = &
 \frac{1}{\cal{N}}  
 \sum_{\{{\vec{\sigma}}_x\}}{e^{\beta S + h_r M} \sin{(h_i M)}}
\, ,  
\nonumber
\\
 ~
 & = &
 Z_L(\beta,h_r)
 \langle
 \sin{(h_i M)}
 \rangle_{\beta,h_r}
\label{ImZ}
\, .
\end{eqnarray}
According to the Lee--Yang theorem the zeroes are on the imaginary
$h$--axis ($h_r = 0$) where
$ \langle \sin{(h_i M)} \rangle_{\beta,h_r}$ vanishes and so the 
Lee--Yang zeroes are simply the zeroes of
\begin{equation}
 {\rm{Re}} Z_L(\beta,i h_i)
 \propto
  \langle
 \cos{(h_i M)}
 \rangle_{\beta,0}
\, .
\end{equation}
Thus the Lee--Yang zeroes are easily  found. Moreover, one can
find them at any $\beta$ using reweighting techniques \cite{histo}.
For  the  lowest  lying  zeroes at 
$\beta = 1.11$  we found   $\theta_1(L) = 0.0023353(7)$,    $0.0006350(2)$,
$0.00017278(5)$ and $0.000047062(13)$  for $L=32,64,128$ and $256$
respectively. 
Errors were calculated using the bootstrap method where 
the data for each $\beta_o$ are resampled 100 times
(with replacement) leading to 100 estimates for $\theta_1$,
from which the variance and bias can be calculated.

\begin{figure}[htb]
\vspace{9pt}
\vspace{7cm}
\includegraphics{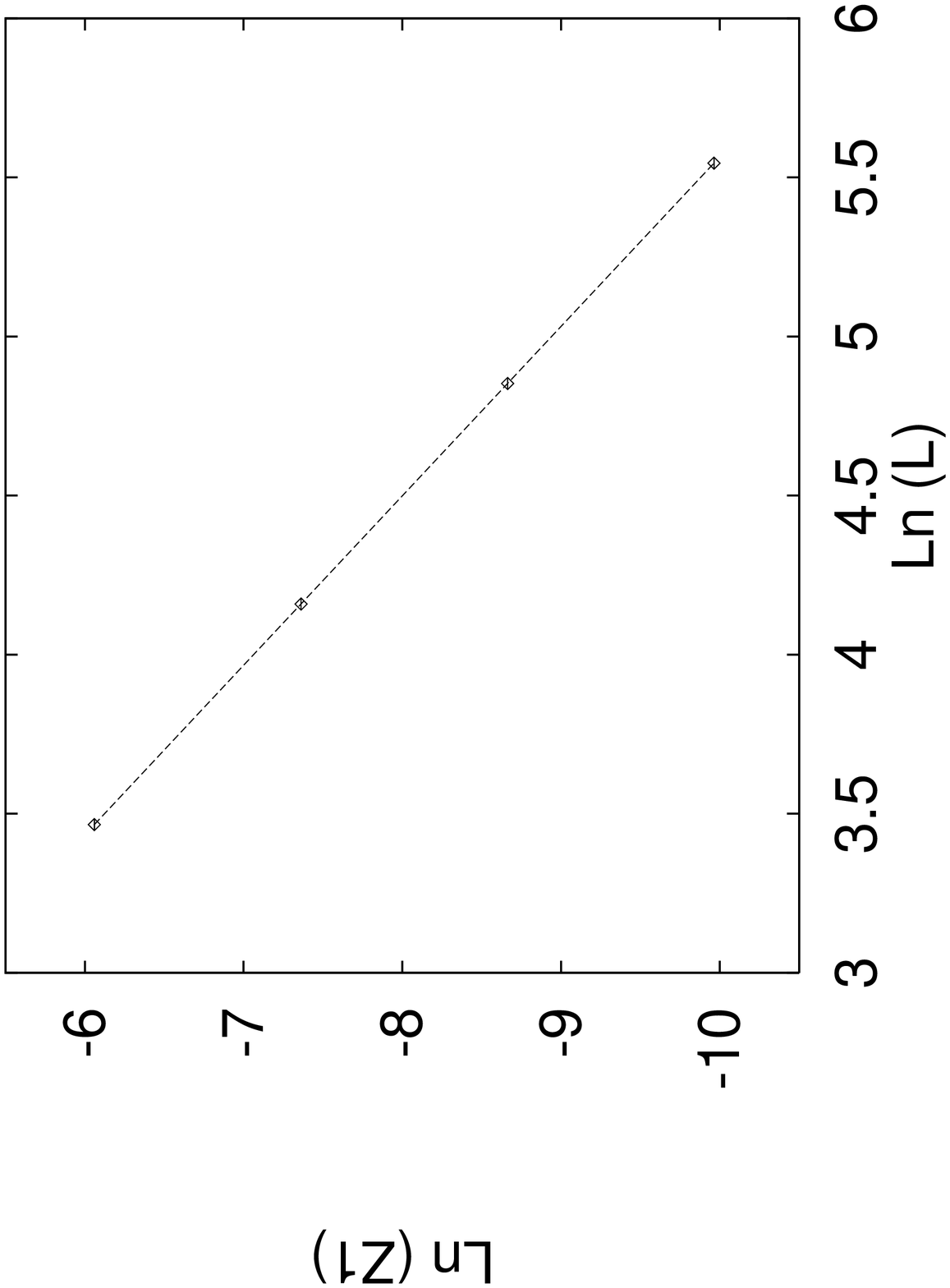}
\caption{Leading FSS of Lee--Yang Zeroes at $\beta = 1.11$.}
\label{fig:leading}
%
\vspace{10.5cm}
\includegraphics{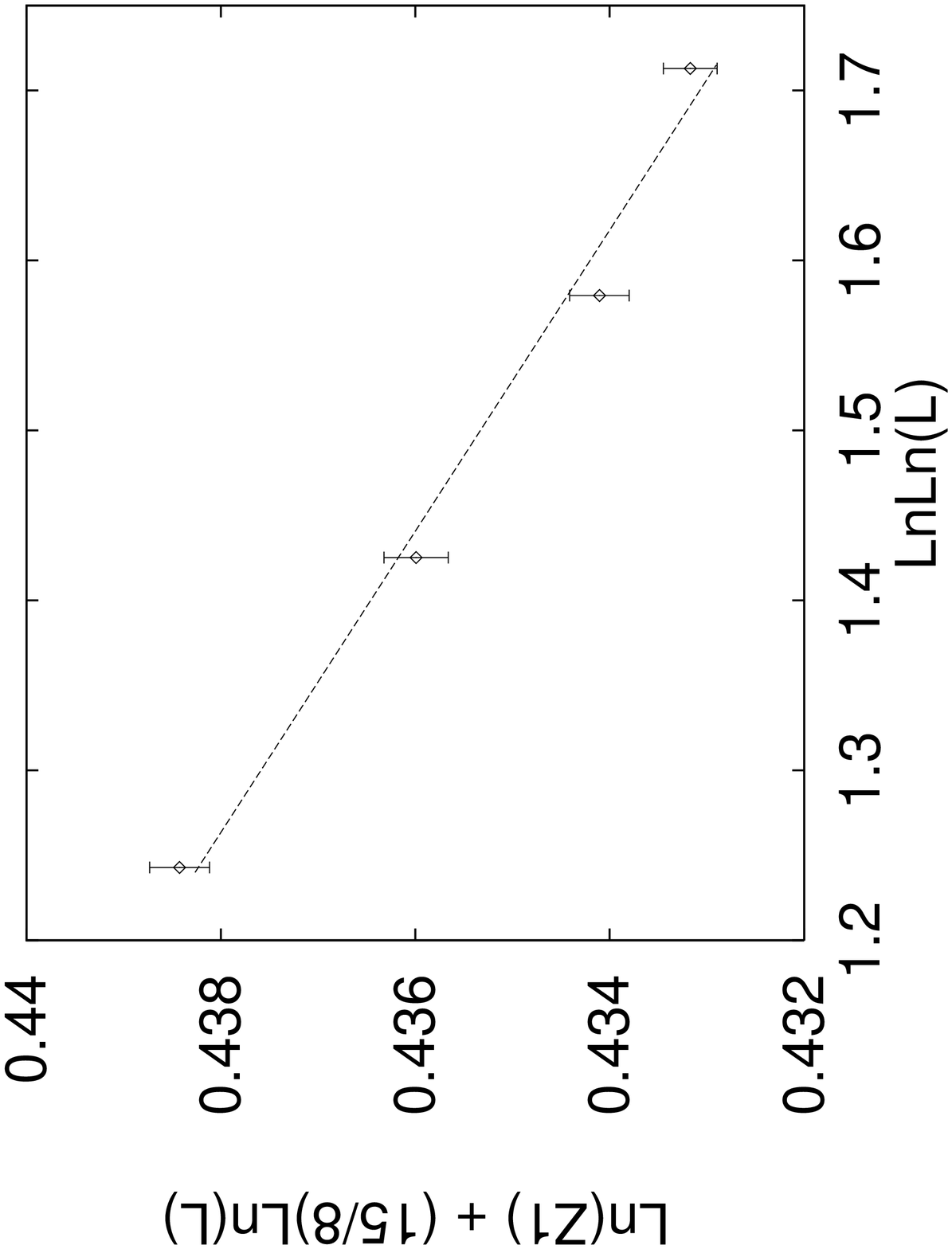}
\caption{Corrections to FSS of Lee--Yang Zeroes at $\beta = 1.11$.}
\label{fig:corrections}
\end{figure}

In figure~\ref{fig:leading} we plot the logarithm of the position of the 
first Lee--Yang zero against the logarithm of the lattice size $L$, at
$\beta=1.11$.
In the absence of any corrections, the slope 
should give the leading power--law  exponent $\lambda$.    
In fact the slope is 
-1.8776(2),  the deviation from the KT value of $-15/8 = -1.875$   being 
attributed to the presence of logarithmic corrections. To identify  
these,  and 
the correction exponent $r$ in  (\ref{goal}),  
$\ln{(\theta_1 L^{15/8})}$  is 
plotted   against  $\ln{\ln{L}}$   in    figure~\ref{fig:corrections}. A 
straight line is identified.   Its  slope  is $-0.012(1)$ 
giving evidence for a non--zero value of 
$r$,   albeit   not  in   agreement  with the RG predictions of 
$-1/16 = -0.0625$ from \cite{KT}.

\begin{figure}[htb]
\vspace{9pt}
\vspace{7cm}
\includegraphics{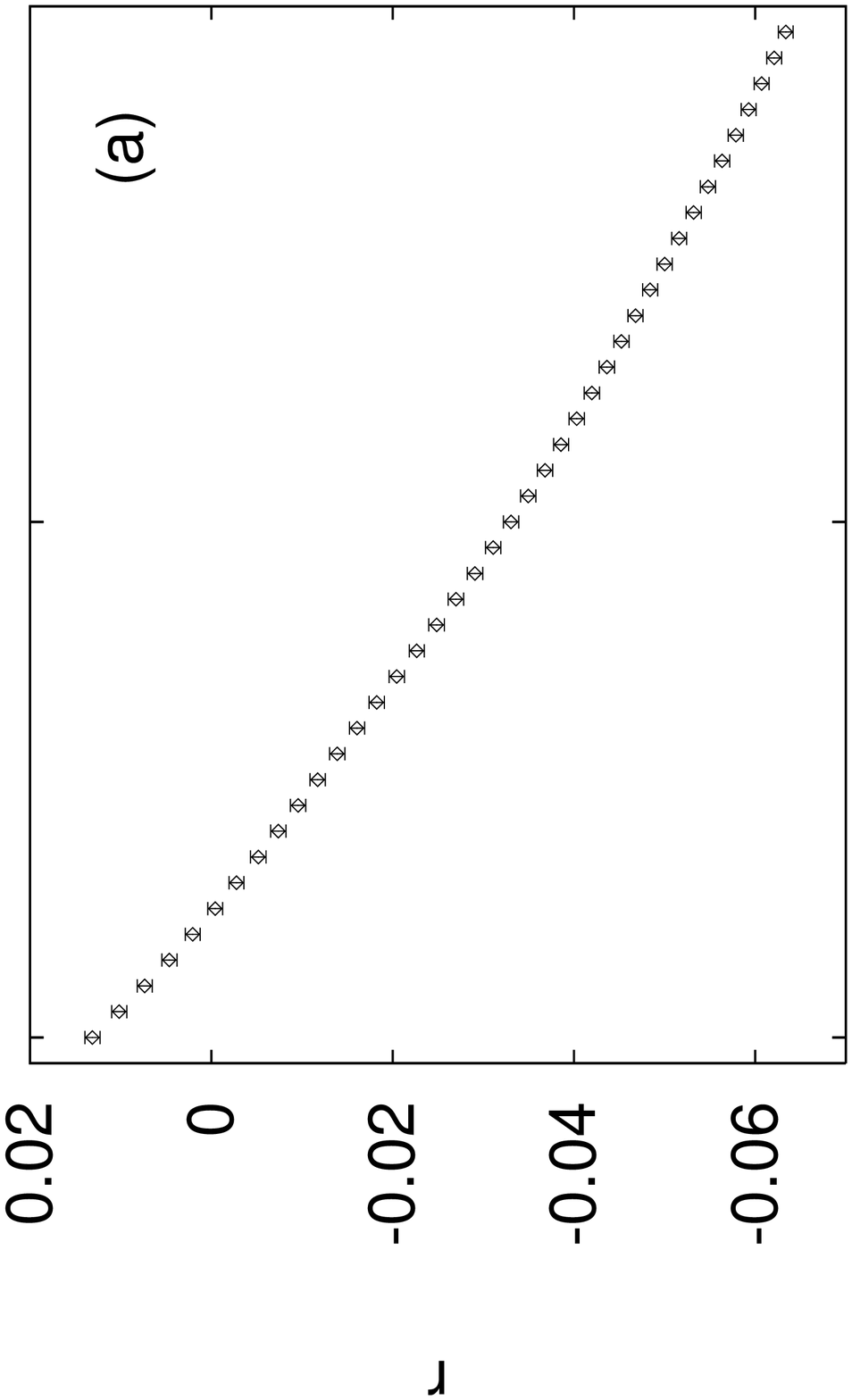}
%
\vspace{6.75cm}
\includegraphics{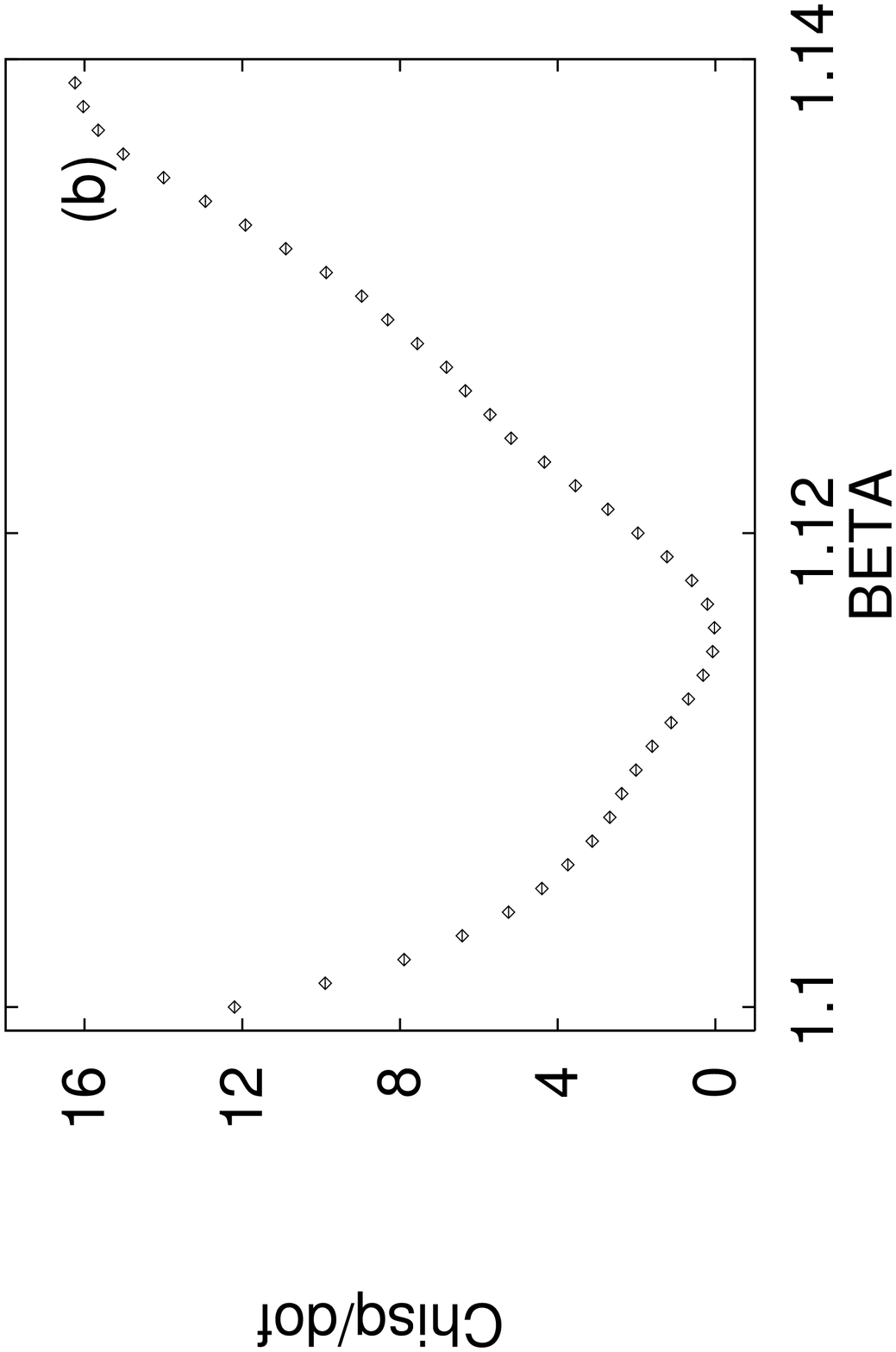}
\caption{(a) Correction Exponent $r$ Measured over a Range of $\beta$
and (b) corresponding $\chi^2$ per Degree of Freedom (Goodness of Fit).}
\label{fig:rbeta}
\end{figure}

At this point, one must investigate the extent to which these 
conclusions regarding
the existence of multiplicative logarithmic corrections depend upon 
the measurement of $\beta_c$. We have attempted to do this
systematically by studying the above $\ln{\ln{L}}$ fits for $r$ 
as a function of the assumed critical beta. 
Fits with a range of
values of $r$ (including zero) are possible but not all with acceptable
$\chi^2$/degree of freedom. 
To be conservative, we take \lq acceptable\rq{} to mean 
less than 2.
This corresponds to
a minimum confidence level of 14\%.
In  figure~\ref{fig:rbeta} 
we show that acceptable
values of $\chi^2$ 
are possible only for 
$1.110$ \raisebox{-.75ex}{ {\small \shortstack{$<$ \\ $\sim$}} } 
$\beta_c$
\raisebox{-.75ex}{ {\small \shortstack{$<$ \\ $\sim$}} } $1.120$.  
The $r=0$ solution would correspond to $\beta_c \approx 1.105$ and
$\chi^2/dof \approx 4 $. 
Similary, we find that a fit with $r=-1/16$
would correspond to $\beta_c \approx 1.138$ and 
$\chi^2/dof \approx 16$. 

In summmary, assumming the Kosteritz-Thouless value $\eta=1/4$, we 
find 
non-zero logarithmic corrections to scaling and a corresponding
estimate of the critical temperature:
\beq
r=-0.023 \pm 0.010 \, ,\qquad\beta_c=1.115\pm.005\, . 
\label{eq:results}
\eeq

As a cross check, one may use the above values for $r$ and $\eta$ 
to find the Roomany-Wyld approximant (\ref{eq:RWA}) and estimate
$\beta_c$ from its zeroes. We find $\beta_c=1.115 \pm 0.004$ 
in good agreement
with the above. However, we emphasize the much higher precision
available to an analysis based on Lee-Yang zeroes rather than on
the spin susceptibility.

\paragraph{Conclusions}
 

Theoretical arguments concerning the consistency of the scaling
behaviour of odd and even  thermodynamic  functions  at  a KT
phase transition have  been  presented.  The  generally  used
scaling  formulae  have  to  be  modified  by  multiplicative
corrections.   These  are  identified  analytically  for  the 
specific heat and numerically for the   susceptibility.  This 
numerical identification comes via an analysis  of  Lee--Yang 
zeroes,  the   FSS  of  which  is   linked  to  that  of  the 
susceptibility.   \\

We would like to thank P. Lacock for assistance with multihistogram
reweighting techniques.

\newpage 


\newpage

\begin{thebibliography}{1234567}
\newcommand{\bibi}[1]{\bibitem{#1}}
\newcommand{\authors}[1]{#1, }
\newcommand{\journal}[1]{#1}
\newcommand{\volume}[1]{{\bf #1}}
\newcommand{\myyear}[1]{(#1)}
\newcommand{\page}[1]{#1}
\newcommand{\mytitle}[1]{}
\newcommand{\keywords}[1]{}
\newcommand{\kw}[1]{}

\bibi{KT}    
\authors{J. Kosterlitz and D. Thouless}           
\journal{J. Phys. C}           
\volume{6} \myyear{1973} \page{1181};
\authors{J. Kosterlitz}
\journal{J. Phys. C}
\volume{7} \myyear{1974} \page{1046}.
 
\bibitem{LY}
C.~N. Yang and T.~D. Lee, Phys. Rev. {\bf 87}  404  (1952);
ibid. 410.

\bibi{DuNe75}   
\authors{F. Dunlop and C.M. Newman}
\journal{Commun. Math. Phys.}
\volume{44} \myyear{1975} \page{223}.
 
\bibi{KeLa94}
\authors{R. Kenna and C.B. Lang}
\journal{Phys. Rev. E}
\volume{49} \myyear{1994} \page{5012}.

\bibitem{Salmhofer}
M. Salmhofer, UBC--Preprint (unpublished);
Nucl. Phys. B (Proc. Suppl.) {\bf 30}, (1993), 81.
 
\bibitem{Abe}
R. Abe, Prog. Theor. Phys. {\bf 37} (1967) 1070;
Prog. Theor. Phys. {\bf 38} (1967) 72; ibid. 322; ibid. 568.

\bibitem{SuzukiLY}
M. Suzuki, Prog. Theor. Phys. {\bf 38} (1967) 289; ibid. 1225;
ibid. 1243;
Prog. Theor. Phys. {\bf 39},  (1968) 349.
 
\bibitem{Fi72}      M.E. Fisher, in Critical Phenomena, Proc.
                    51th Enrico Fermi Summer School, Varena, ed. M.S. Green
                    (Academic Press, NY, 1972).

\bibi{KeLa93}
\authors{R. Kenna and C.B. Lang} 
\journal{Phys. Lett. B}
\volume{264} \myyear{1991} \page{396};
\journal{Nucl. Phys. B (Proc. Suppl.)}
\volume{30} \myyear{1993} \page{697};
\journal{Nucl. Phys. B}
\volume{393} \myyear{1993} \page{461};
Err. ibid.  B {\bf 411} (1994) 340.

\bibi{Lu82}
\authors{J.M. Luck}
\journal{J. Phys. A}
\volume{15} \myyear{1982} \page{L169}.
 
\bibi{IPZ}   
\authors{C. Itzykson, R.~B. Pearson, and J.~B. Zuber}
\journal{Nucl. Phys. B}
\volume{220}[FS8] \myyear{1983} \page{415};
C. Itzykson and J.~M. Luck, 
Progress in Physics, Critical Phenomena 
(1983 Brasov Conference) Birkh{\"a}user, 
Boston Inc. {\bf Vol 11},  45  (1985).

\bibi{BuCo92}
\authors{P. Butera and M. Comi}
\journal{Phys. Rev. B}
\volume{47} \myyear{1993} \page{11969}.

\bibi{AlHa88}
\authors{C. Allton and C. Hamer}
\journal{J. Phys. A}
\volume{21} \myyear{1988} \page{2417}.

\bibi{Wolff}
\authors{U. Wolff}
\journal{Phys. Rev. Lett.}
\volume{62} \myyear{1989} \page{361}.

\bibitem{histo} 
\authors{K. Kajantie, L. K\"arkk\"ainen and K. Rummukainen}
\journal{Nucl. Phys. B}
\volume{357} \myyear{1991} \page{693}.


\bibitem{HI} A.C. Irving and C. Hamer, 
             Nucl. Phys. B {\bf{230}}[FS10] (1984) 361;
C. Hamer and A.C. Irving, 
             J. Phys. A {\bf{17}} (1984) 1649.


\bibitem{RoWy}      M.P. Nightingale, Physica A {\bf{83}} (1976) 561;
                    H. Roomany and H.W. Wyld, 
                    Phys. Rev. D {\bf{21}} (1980) 3341.

 
  
\bibi{many}       
\authors{R. Gupta, J. De Lapp, G. Batrouni, G.C. Fox, C.F. Baillie
and J. Apostolakis}
\journal{Phys. Rev. Lett.}
\volume{61} \myyear{1988} \page{1996};
\authors{L. Biferale and R. Petronzio}
\journal{Nucl. Phys. B}
\volume{328} \myyear{1989} \page{677}; 
\authors{C.F. Baillie and R. Gupta} 
\journal{Nucl. Phys. B (Proc. Suppl.)} 
\volume{20} \myyear{1991} \page{669};
\authors{A. Hulsebos, J. smit and J.C. Vink}
\journal{Nucl. Phys. B}
\volume{356} \myyear{1991} \page{775}; 
\authors{R.G. Edwards, J. Goodman and A.D. Sokal}
\journal{Nucl. Phys. B}
\volume {354} \myyear{1991} \page{289}.
 

\bibi{HaMa94} 
\authors{M. Hasenbusch, M. Marcu and K. Pinn}
\journal{Physica A}
\volume{208} \myyear{1994} \page{124}.

 
 

\end{thebibliography}
\end{document}